\documentclass[letter]{aa}  

\usepackage{graphicx}
\usepackage{txfonts}
\usepackage{xcolor}
\usepackage{amsbsy,amsmath}
\usepackage{amsfonts}
\usepackage{natbib}
\usepackage[colorlinks,allcolors=blue]{hyperref}

\begin{document}

   \title{Dynamical constraints on planet engulfment as the origin of lithium enhancement in TOI-5882
}

%

   \author{C. Charalambous\inst{1}
   \fnmsep\thanks{\email{ccharalambous@uc.cl}}
          \and
          C. Aguilera-Gómez\inst{1}\fnmsep\thanks{\email{craguile@uc.cl}}
          \and
          S. Urrutia\inst{1}
          \and
          C. Norambuena\inst{1}
        }

   \institute{Instituto de Astrofísica, Pontificia Universidad Católica de Chile, Av. Vicuña Mackenna 4860, 782-0436 Macul, Santiago, Chile}

   \date{Received September 30, 20XX}

 
  \abstract
    {As stars evolve off the main sequence, changes in stellar structure can alter the dynamical architecture of planetary systems and, in some cases, lead to planet engulfment events capable of producing observable chemical signatures such as lithium enhancement. TOI-5882 is a lithium-rich subgiant hosting a $\sim$22 $m_{\rm Jup}$ brown dwarf on a 7.1-day orbit, where the enrichment could plausibly result from the recent engulfment of a super-Earth to Neptune-mass planet.}
   {We assess the dynamical viability of a planet engulfment scenario in the TOI-5882 system that could explain its observed Li enrichment.}
    {We combine stellar evolution models with N-body simulations that incorporate time-dependent stellar properties as the host star leaves the main sequence, exploring a broad range of pre-engulfment planetary masses and orbital configurations.}
    {Planet engulfment is the most probable outcome under the explored configurations. However, only 5\% of the simulations produce engulfment within the short detectability window of the lithium-enrichment signature. Engulfment therefore remains a viable explanation for the observed Li enhancement in TOI-5882, but only for a relatively uncommon subset of the initial conditions considered here.}
   {Under our simplified architecture, recent engulfment cannot be ruled out as the origin of the observed lithium enhancement. However, because successful cases represent only a small fraction of the initial conditions explored here, and because additional processes such as wind-induced drag forces were not included, additional constraints on the system's primordial architecture are needed to assess how likely this scenario is for TOI-5882. }


   \maketitle

\section{Introduction}

The fate of planets and their host stars is intrinsically linked. While they interact dynamically from birth, stellar evolution introduces additional effects by altering the star's radius, mass, and luminosity.
After the main sequence, planets in tight orbits may be engulfed as the star expands along the red giant branch \citep[RGB,][]{2009ApJ...705L..81V,2020ApJ...898L..23R} and asymptotic giant branch \citep[AGB,][]{2012ApJ...761..121M} phases, potentially producing observable signatures of these interactions, such as enhanced light element abundances \citep{2016ApJ...829..127A, AG2020} and increased stellar rotation \citep{carlberg2012,privitera2016}.

TOI-5882 is a subgiant star that shows high lithium (Li) abundances compared to other stars in the same stage of evolution, with similar mass and metallicity, and hosts a $\sim 22 \, m_{\rm Jup}$ brown dwarf (BD) on a 7.1-day slightly eccentric ($e_{\rm BD}=0.0339$) orbit \citep{vowell2025}. Given that this star is located in the most favourable window within the subgiant branch to retain detectable Li enrichment following planet engulfment, according to evolutionary models \citep{2021AJ....162..273S}, \cite{2026arXiv260521407K} propose that the observed abundances may be explained by the deposition of planetary material into the stellar convective envelope, and they provide an estimate of the required accreted mass. Using estimates of the planetary Li abundance of \cite{2016ApJ...829..127A}, they further suggest that TOI-5882's Li enhancement could plausibly result from the ingestion of a super-Earth to Neptune-mass planet ($\sim 9-95\, m_\oplus$).

\cite{2026arXiv260521407K} also propose that the presence of the companion brown dwarf may have driven dynamical perturbations leading to the engulfment event, which is not inferred from the high stellar Li abundance. This scenario motivated our dynamical investigation.

In this work, we perform N-body simulations that incorporate the star's evolution as it leaves the main sequence to explore a broad range of possible pre-engulfment planetary masses and orbital periods. Our goal is to assess whether planet engulfment is expected to occur within the evolutionary time window required to produce an observable Li signature, and therefore, to determine whether the engulfment scenario can plausibly account for the high Li abundance observed in TOI-5882.

Our study investigates the dynamical viability of a planet engulfment scenario in the TOI-5882 system by numerically exploring a physically motivated range of parameters for a putative planet that may have been part of the system in the past. By assessing whether such an engulfment event can occur within the relevant time window, we evaluate if this scenario could account for the high Li abundance observed in the host star. The paper is organized as follows: in Section \ref{sect:methods} we describe the numerical setup and initial conditions of our simulations, in Section \ref{sect:resultados} we present our main results and in Section \ref{sect:discusion} we discuss the interpretation of the engulfment fraction in TOI-5882. Finally, in Section \ref{sect:conclusions} we summarize our conclusions.


\section{Methods} \label{sect:methods}

\begin{figure}
    \centering
    \includegraphics[width=\columnwidth]{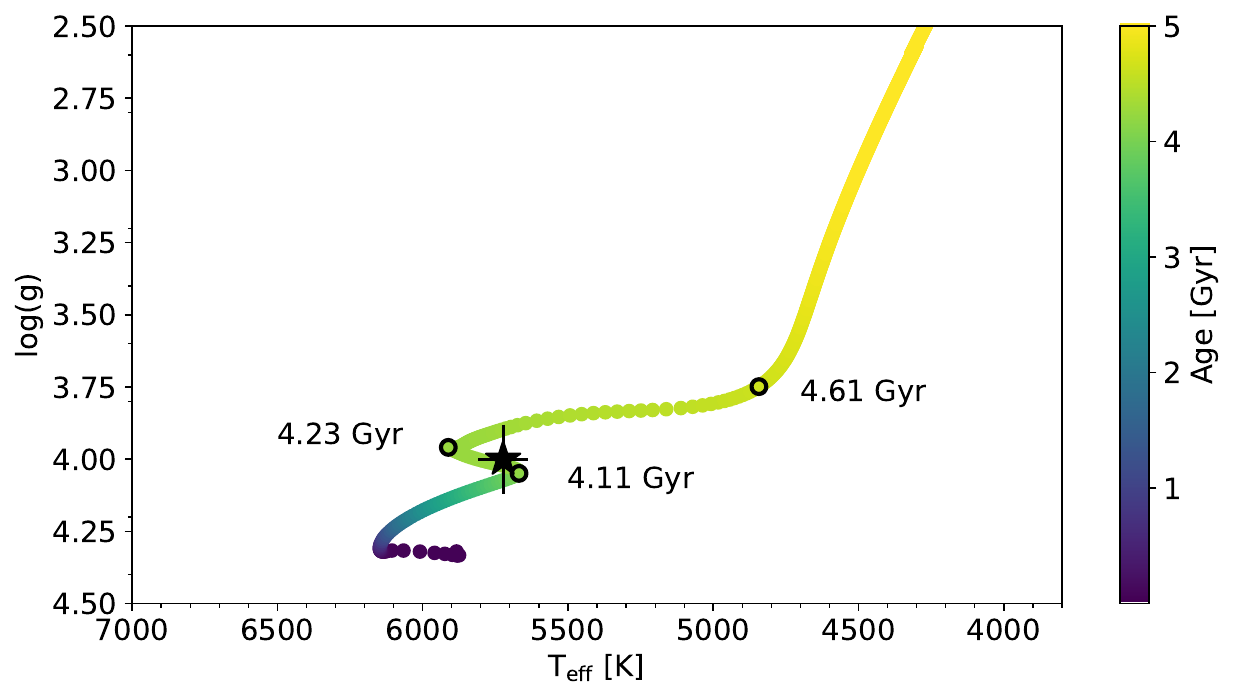} \\
    \includegraphics[width=\columnwidth]{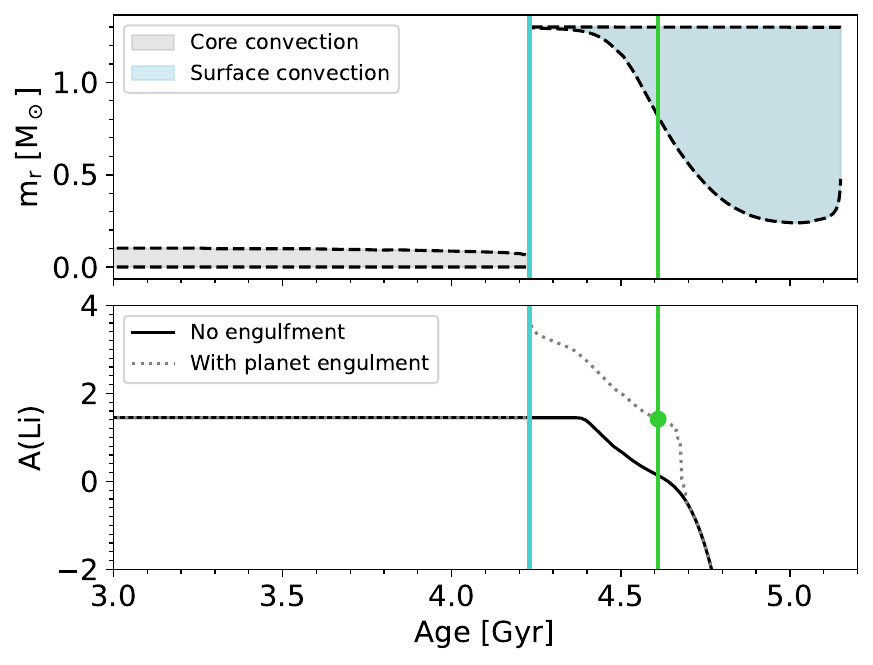} 
    \caption{\textbf{Top:} $T_{\mathrm{eff}}$-$\log g$ diagram showing the evolutionary track of a $1.3\ \mathrm{M_\odot}$ star with $[\mathrm{Fe/H}] = 0.33$ dex, color-coded by stellar age. The position of TOI-5882 is indicated using the atmospheric parameters reported by \cite{2026arXiv260521407K}. With circles we highlight the age reported in previous studies (4.11 Gyr); the onset of the subgiant branch (4.23 Gyr); and the end of the evolutionary phase during which surface Li remains detectable (4.61 Gyr). \textbf{Middle:} Kippenhahn diagram showing the evolution of the convective zone (shaded regions). Vertical lines indicate the time interval explored in the simulations. \textbf{Bottom:} Time evolution of surface Li abundance, A(Li), considering the baseline value reported by \cite{2026arXiv260521407K} (solid line) and the abundance after planet engulfment (dotted gray line). The green point marks the time at which post-engulfment Li becomes observationally indistinguishable from the baseline value.}
    \label{fig:HRage}
\end{figure}
 
Given the nature of the problem, our methodology combines stellar-evolution calculations with direct N-body integrations, allowing us to model the dynamical evolution of a planetary system orbiting a star whose properties vary over time.

To model the post–main-sequence evolution of the star, we use MESA \citep[Modules for Experiments in Stellar Astrophysics,][version r24.08.1]{2011ApJS..192....3P}.
We compute stellar evolutionary models assuming a non-rotating star with a metallicity of $[{\rm Fe/H}] = 0.33$ dex. and adopt an initial stellar mass of $M_\star =1.3 \, M_\odot$, fully consistent with the spectroscopic constraints reported by \cite{2026arXiv260521407K}. With these parameters, the stellar evolutionary track places TOI-5882 near the subgiant branch, as shown in the top panel of Figure \ref{fig:HRage}. 
Uncertainties in stellar mass and atmospheric parameters naturally account for the small offset between the best-fit track and the position of the star in the $T_{\mathrm{eff}}$–$\log g$ diagram, such that consistency with the early subgiant branch is achieved once the reported errors are taken into account (see Appendix \ref{app:st_params} for details).

For the N-body integrations, we use the open-source code REBOUND \citep{2012A&A...537A.128R} to simulate the dynamical evolution of planets orbiting an evolving central star. For our purposes, we use the high-accuracy IAS15 integrator \citep{2015MNRAS.446.1424R} that employs adaptive timesteps and properly handles close encounters either between the planet and the brown dwarf or with the star during engulfments.

Because stellar properties evolve on gigayear timescales while the orbits of the close-in brown dwarf and putative inner planet vary orders of magnitude faster, stellar parameters change slowly relative to the orbital motion. Recomputing a full stellar model at each N-body timestep would therefore be computationally prohibitive. Instead, we compute a single MESA stellar-evolution track and interpolate its output across a large ensemble of dynamical simulations, using MESA's outputs as inputs for REBOUND.

We incorporate evolving stellar properties with the REBOUNDx library \citep{2020MNRAS.491.2885T}, using its parameter-interpolation module \citep{2022MNRAS.510.6001B} to ensure a smooth coupling between the stellar-evolution track and the N-body dynamics. From the MESA models, we extract the time-dependent stellar mass, radius, and luminosity. These quantities are used to compute the tidal forces within REBOUNDx using the equilibrium-tide constant time-lag model \citep{1981A&A....99..126H}. Tidal interactions are included for both the putative planet and the BD, with the stellar tidal Love number computed following \citet{2013ApJ...778..100B}, allowing it to vary with the properties of the stellar convective envelope throughout the evolution.
\begin{figure}
    \centering    \includegraphics[width=\columnwidth]{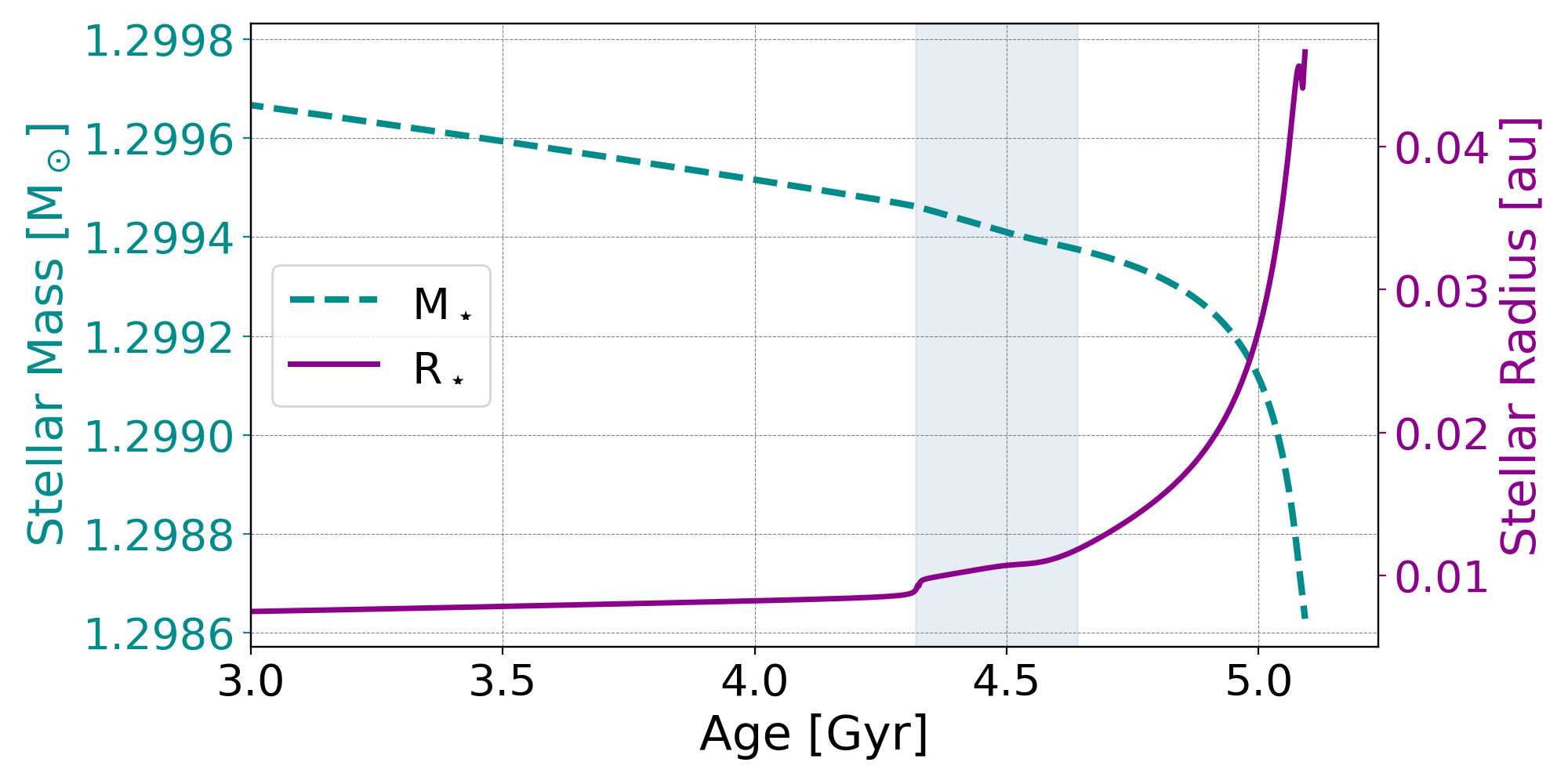}
    \caption{Temporal evolution of the stellar mass (dashed line) and radius (solid line) from the end of the main sequence through the completion of the subgiant phase. The gray shaded area indicates the time interval during which a planetary engulfment would be observationally detectable. }
    \label{fig:TempEvol}
\end{figure}

During the giant-branch phases, the dominant tidal effect arises from tides raised by the companions on the extended convective envelope of the host star, which can drive orbital decay and modify the minimum orbital distance required for engulfment \citep[e.g.,][]{2007ApJ...661.1192V,2011ApJ...737...66K,2012ApJ...761..121M,2016RSOS....350571V}. By contrast, tides raised on the companions are expected to play a comparatively minor role in the post-MS evolution, although they are included self-consistently in all integrations. We note, however, that tidal dissipation in evolved stars remains uncertain, and the adopted prescription may under- or overestimate the true strength of tidal interactions \citep{2012ApJ...761..121M}.

We thus consider a system consisting of an evolving $1.3\,M_\odot$ star, followed from the late main-sequence phase through the end of the subgiant branch. The system hosts a $22\,m_{\rm Jup}$ BD, initialized with its observed orbital configuration ($P_{\rm BD}=7.1$~days, $e_{\rm BD}=0.0339$). In addition, we include an inner planet whose mass, orbital period, eccentricity, and inclination are drawn randomly from the ranges $[9,95]\,m_\oplus$, $[0.5,3.5]$~days, $[10^{-4},10^{-3}]$, and $[-1^\circ,1^\circ]$, respectively.
Our approach is agnostic to the formation history of the system. Rather than modeling their origin, we focus on testing the proposed engulfment hypothesis and assessing the dynamical conditions under which such configurations can lead to planetary engulfment during stellar evolution, without making assumptions about the formation mechanisms of the planetary or substellar companions.

Having defined the numerical setup and stellar models, we now examine the evolutionary time window over which a planet engulfment event could lead to an observable Li abundance.
\begin{figure*}
    \centering
    \includegraphics[width=\textwidth]{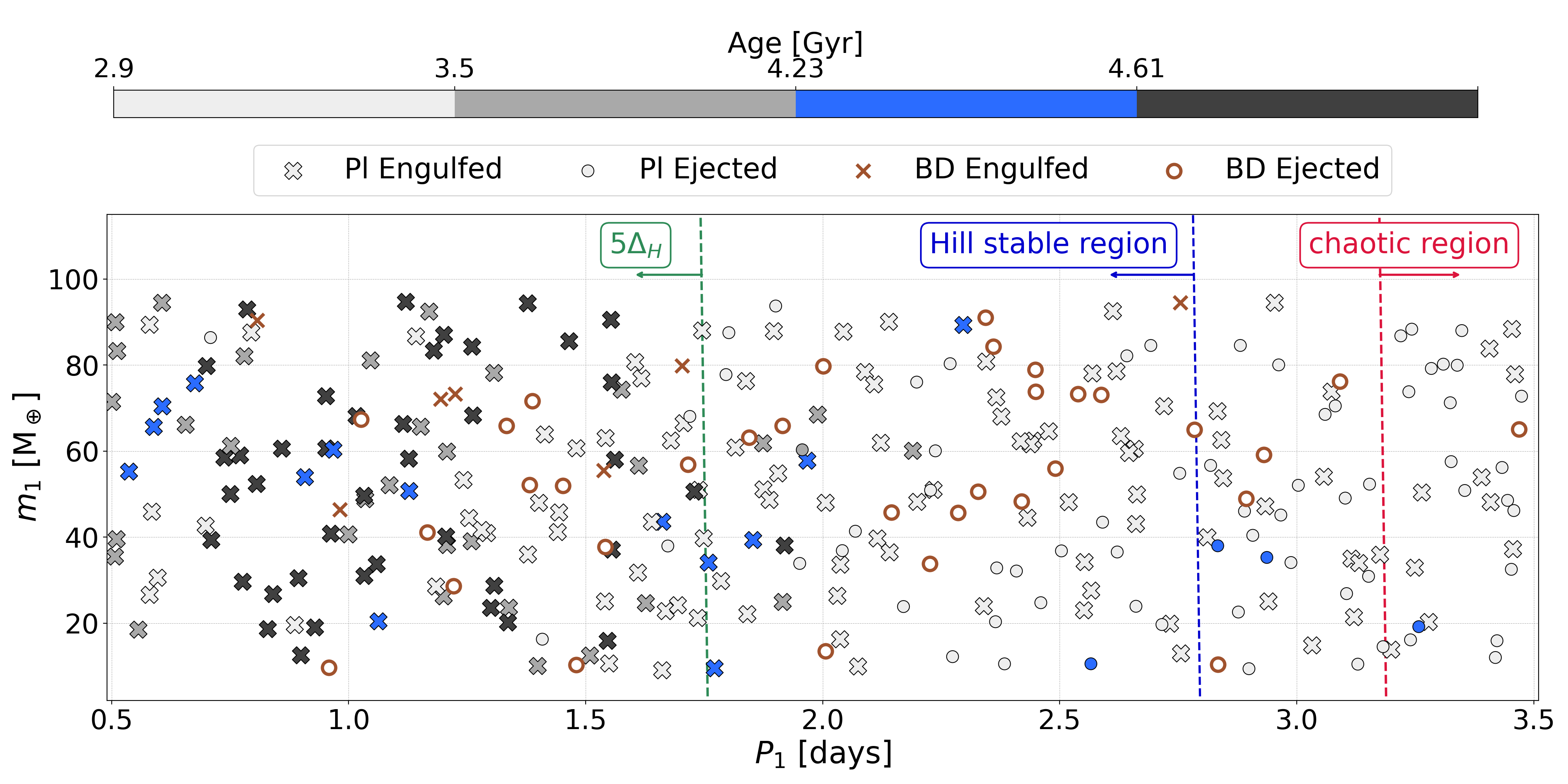}
    \caption{Outcome of 300 simulations exploring plausible pre-engulfment orbital configurations in the period–mass plane. The blue dashed line mark the common stability threshold mutual Hill-radius separation between a fixed brown dwarf ($m=22\,m_{\rm Jup}, P_{\rm BD}=7.1\,$d), and an inner planet with varying mass, around a $M_\star=1.3\,M_\odot$ star. The green dashed line represent a more conservative stability limit.  Crosses mark cases in which the planet is engulfed by the evolving star, while filled circles indicate planetary ejections, color-coded by the time at which the event occurs. Brown open circles and crosses denote BD escapes and engulfments.}
    \label{fig:sabtility}
\end{figure*}

\section{Results} \label{sect:resultados}

The stellar evolutionary models allow us to define the temporal window during which accreted planetary material can be observationally identified through Li enhancement. Given the intrinsic ambiguity between stellar models (Appendix \ref{app:st_params}), we do not adopt a single age for TOI-5882. Instead, we define an evolutionary time window directly from the stellar models during which a planet engulfment event could produce observable signatures. This window, indicated in the $T_{\mathrm{eff}}$–$\log g$ diagram in Fig. \ref{fig:HRage}, corresponds to the onset of the subgiant branch and the development of a surface convective envelope, as illustrated in the Kippenhahn diagram (middle panel) and discussed by \citet{2021AJ....162..273S}.

The temporal boundaries of our simulations are set by the physical considerations related to stellar structure and Li observability. In our MESA models, this phase begins at 4.23 Gyr (vertical cyan line in the middle and bottom panel of Fig. \ref{fig:HRage}), when accreted planetary material can be efficiently mixed into the surface convective envelope. We therefore consider only epochs for which the post-engulfment Li abundance exceeds the baseline value of $A(\mathrm{Li}) = 1.45$ dex reported by \cite{2026arXiv260521407K} based on GALAH DR4 data \citep{2025PASA...42...51B} for comparable stars. Once the surface Li abundance drops below this level, any planetary contribution becomes observationally indistinguishable.

The bottom panel of Fig. \ref{fig:HRage} shows the evolution of surface Li abundance with and without planet engulfment (see Appendix \ref{app:st_params} for details). The green point marks the end of the detectable phase at 4.61 Gyr (vertical green line).\footnote{Notice that this time interval is broader than that considered by \cite{2021AJ....162..273S}, who focused on the very beginning of the subgiant branch to detect, with high significance, the engulfment of a $1\,m_{\rm Jup}$ planet.} Over this relatively short window, stellar mass and radius vary only modestly, with minimal impact on the planetary dynamics (Fig. \ref{fig:TempEvol}). In addition to dilution from the first dredge-up, Fig. \ref{fig:HRage} shows that Li is efficiently destroyed in the convective envelope leading to a rapid decrease in surface A(Li). Li delivered by an engulfed planet is affected by the same burning, imposing strong timing constraints on any viable engulfment scenario.

According to \citet{2026arXiv260521407K}, explaining the observed Li abundance in the central star requires the ingestion of a planet with a mass between 9 and 95 $m_\oplus$ (see Appendix \ref{app:pl_params} for a discussion on planet parameters). To test this hypothesis, we performed 300 simulations with randomly sampled planetary masses and orbital periods.
We evolve each simulation until the host star enters the giant phase or until either the planet or the BD is removed from the system through escape or stellar engulfment, whichever occurs first. When the BD is lost we terminate the simulation, as it constitutes the only currently observable companion in the system. We record the time at which any event affecting the planet occurs, in order to assess whether it falls within the narrow temporal window during which Li abundance is observationally detectable in TOI-5882.

Fig.~\ref{fig:sabtility}  summarizes the outcomes of the 300 simulations performed in this study. Brown open circles and crosses denote scenarios in which the brown dwarf escapes the system or is engulfed by the star, respectively. Filled crosses and circles indicate cases in which the planet is engulfed by the evolving host star or ejected from the system, color-coded according to the time at which the event occurs. 
The orbital period of the inner planet is chosen such that its initial separation from the BD spans both dynamically stable and unstable configurations (see Appendix \ref{app:pl_params} for details of the stability boundaries).

For planetary orbital periods $P\gtrsim2$ days, the majority of dynamical events—both ejections and engulfments—occur before $\sim$3.5 Gyr (light gray). Systems in which the planet is initially placed closer to the BD are markedly more unstable, typically leading to rapid planetary ejection at early times.

As the initial configuration approaches the Hill-stable regime, a mixture of planetary engulfments and ejections is observed, predominantly around the same time. During the post-main-sequence phase, planetary orbits destabilize, mainly due to their gravitational interactions, and not because of the stellar expansion. When the separation between the planet and the brown dwarf exceeds approximately five or six mutual Hill radii, the planet survives in most cases from 3.5 Gyr up to nearly 4 Gyr (in gray).

\begin{figure}
    \centering
    \includegraphics[width=\columnwidth]{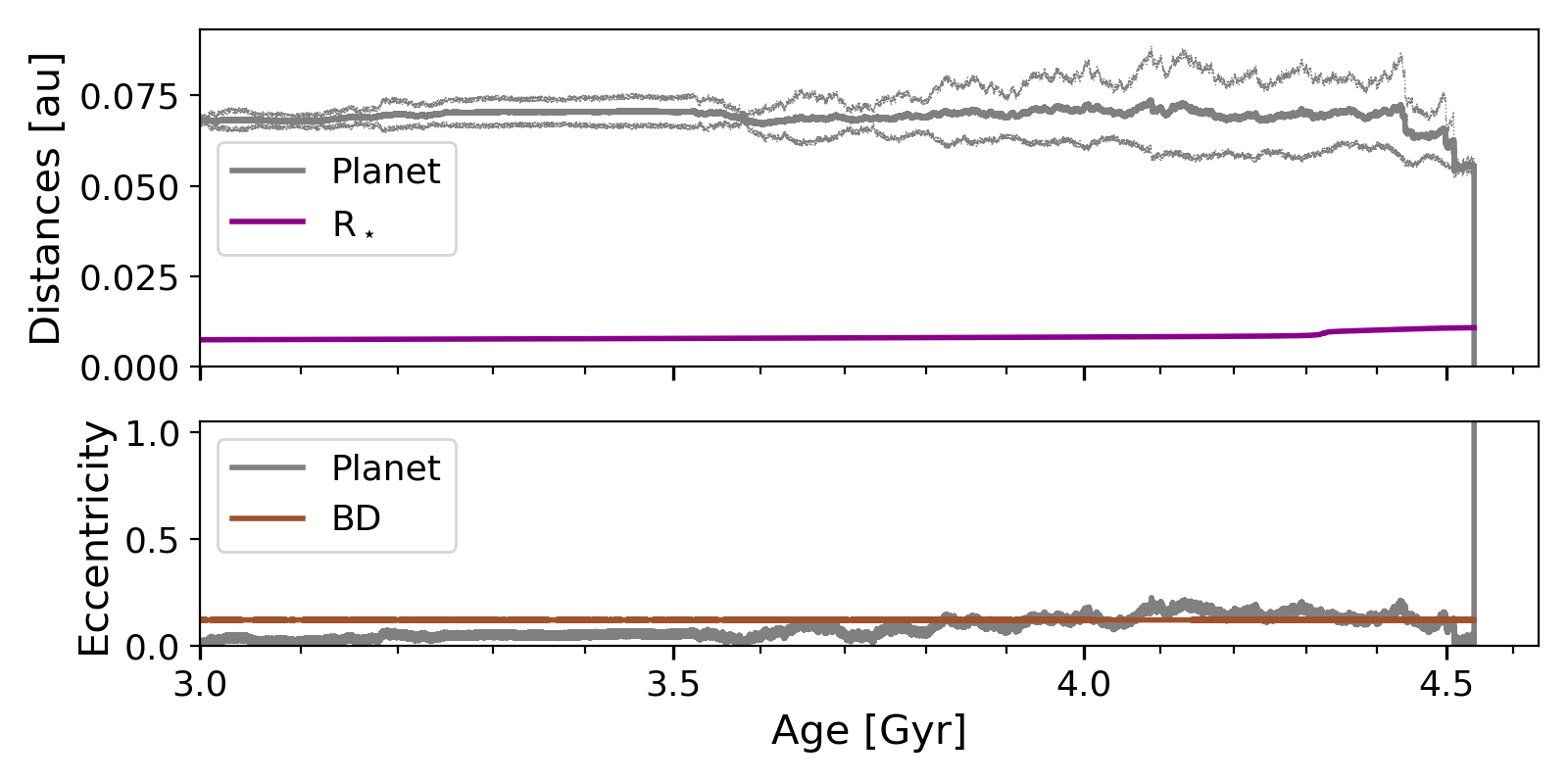}
    \caption{Planet engulfment occurring during the lithium-detectable period ($m_{\rm pl} = 53.95\,m_\oplus$, $P_{\rm pl} = 0.91$\,d).  Evolution of the semi-major axes, pericenter,  apocenter distances (top) and eccentricities (bottom) of both the planet and the BD, together with the stellar radius. }
    \label{fig:aqQe}
\end{figure}

Among all 300 simulations, only less than 5\% result in planetary engulfment during the narrow time interval between 4.23 and 4.61 Gyr, when Li enrichment would be observationally detectable (blue crosses) and associated to the engulfment of the planet. We find no clear trend or correlation either with the mass or with the orbital period of the simulated planet, suggesting that they are instead, sporadic events. An example of such a case is shown in Figure \ref{fig:aqQe}, where the planetary system becomes destabilized and the planet is engulfed at $\sim$4.54 Gyr, while the BD maintains its observed orbital period and eccentricity. The rarity of engulfment events occurring within the Li detection window indicates that this scenario is intrinsically unlikely.

It is important to note that, out of our 300 simulations, approximately 65\% result in planet engulfment, although most of these events do not occur within the time interval during which the observed lithium abundance can be directly associated with the accretion of planetary material. This relevant time window is constrained by the stage at which the engulfed material can be efficiently mixed within the stellar convective envelope.

Nevertheless, about 45\% of the engulfment events take place during earlier evolutionary phases, when the outer stellar layers are radiative. In this regime, the fate of the accreted material, and in particular the evolution of lithium abundance, is considerably more uncertain, making it difficult to assess whether such events could produce an observable signature.

Therefore, planet engulfment emerges as the most probable outcome in our simulations. However, reproducing the observed lithium abundance requires that the engulfment occurs within a relatively narrow evolutionary interval under convective conditions,  making timing a critical factor in assessing whether this mechanism can explain the observations.

Finally, although the Li detectability window ends at 4.61 Gyr, we extend the simulations until either BD or planet are engulfed or escape the system. A higher fraction of engulfment events occur after this epoch. However, once the first dredge-up is advanced, Li is rapidly destroyed under convective conditions, so any accreted planetary material would be depleted on very short timescales and remain undetectable. Such late engulfment events cannot account for the observed Li enrichment in TOI-5882.

\section{Discussion}\label{sect:discusion}

A key point in interpreting our simulations is that the quoted $\sim 5\%$ fraction of recent engulfment events should not be understood as the probability that TOI-5882 itself experienced a recent planet engulfment episode. Rather, it refers to the fraction of the explored initial conditions in which a hypothetical inner planet is engulfed within the observationally relevant time window over which a Li-enrichment signature would remain detectable. Our numerical experiment is therefore intended to test whether the currently observed architecture of TOI-5882 is dynamically compatible with a recent engulfment event, rather than to infer the a posteriori probability that engulfment occurred in this specific system. In this sense, the low fraction of recent engulfment events does not imply that engulfment is implausible for TOI-5882; instead, it indicates that, within the simplified set of initial conditions explored here, recent engulfment is a possible but non-generic outcome.

We note that the orbit of the BD is expected to evolve due to tidal interactions already during the MS (e.g., \citealt{2014ApJ...794....3V}). More recently, tidal calculations tailored specifically to TOI-5882 \citep{2026arXiv260504141N} predict continued orbital decay on a timescale of $\sim$25–125 Myr from the present day, implying that the currently observed post-MS configuration is transient. Incorporating this secular orbital evolution could modify the timescales of scattering and engulfment, potentially bringing some engulfment events into the lithium observational detectability window. Our results should therefore be interpreted as conditional on the currently observed post-MS orbit of the companion.

More generally, our results should not be interpreted as a population-level statement about the origin of Li enhancement in subgiant systems hosting close-in BDs. A robust probabilistic assessment would require a broader study sampling the distribution of companion masses and orbital properties in such systems, together with a comparison between predicted engulfment frequencies and the observed incidence of Li-rich stars. Here, our simulations are intended as a single system motivated dynamical experiment, rather than as a population-level inference on the cause of lithium enrichment. In addition, the present calculations adopt a simplified architecture and do not include other processes that may influence the orbital evolution and engulfment timescales, such as wind-induced drag forces. Incorporating these effects, as well as extending the analysis to a population of BD-hosting systems, would provide a more complete framework for evaluating the role of engulfment in producing Li enhancement in evolved stars.

\section{Conclusions}\label{sect:conclusions}

To model the TOI-5882 Li-rich system, we combine two complementary numerical tools: one accounting for the stellar evolution and another performing the N-body integrations. 

Our main result suggests that, within the engulfment scenario explored here, the timescale over which lithium enrichment remains detectable is short compared to the relevant evolutionary timescales.  This suggests that the currently observed Li abundance in TOI-5882 is compatible with engulfment events occurring within the predicted timescale by a subset of our simulated system configurations. 

More generally, our results suggest that, although a planet engulfment event may still have occurred in TOI-5882, explaining the currently observed lithium abundance through this channel requires a rather specific timing between the engulfment and the present evolutionary state of the star. In this sense, the engulfment scenario is not ruled out, but it appears difficult to reconcile with the observed lithium enhancement under the assumptions adopted here.

It is important to note that our simulations explore a simplified configuration, namely a single planet located interior to the brown dwarf. Alternative architectures could also be considered, such as multiple lower-mass planets whose combined mass matches the estimated value, or configurations in which the planet(s) reside exterior to the brown dwarf. Furthermore, our model does not include other physical processes that may influence the orbital evolution, such as stellar winds or mass-loss–driven effects, which could exert an outward force on a planet undergoing inward migration. Accounting for these additional complexities may alter the detailed evolutionary pathways and will be the subject of future work.

Measuring additional chemical tracers beyond lithium may provide an observational avenue to test the engulfment scenario in this system.

\begin{acknowledgements}
The authors would like to thank the anonymous reviewer for their helpful comments and questions, which helped improve our manuscript. 
CC acknowledges support from Agencia Nacional de Investigación y Desarrollo (ANID) through FONDECYT Postdoctoral grant n$^\circ$3230283 and CAS-ANID (CASSACA) n$^\circ$250005. CAG acknowledges support from ANID through FONDECYT Iniciación 11230741 and FONDECYT Regular 1262342.
The Geryon cluster at the Centro de Astro-Ingenieria UC was used for our simulations. 
ANID BASAL project FB21000, BASAL CATA PFB-06, the Anillo ACT-86, FONDEQUIP AIC-57, and QUIMAL 130008 provided funding for several improvements to the Geryon cluster. The authors would like to thank M. Soares-Furtado for valuable comments and discussions.
\end{acknowledgements}
\bibliographystyle{aa} 
\bibliography{li_bib} 

\begin{appendix}
\nolinenumbers
	
\section{Stellar model caveats} \label{app:st_params}
We based our models on the stellar properties reported in \cite{2026arXiv260521407K}, the effective temperature ($T_{\mathrm{eff}} = 5723 \pm 85$ K), surface gravity ($\log g = 4.0 \pm 0.12$ dex), and metallicity ($[\mathrm{Fe/H}] = 0.33 \pm 0.08$ dex) were derived spectroscopically in that work and provide an internally consistent characterization of the stellar atmosphere. However, the stellar mass ($1.334\,M_\odot$) and age (4.11 Gyr) adopted by \cite{2026arXiv260521407K} were taken from \citet{vowell2025}, who modeled the brown dwarf and host star simultaneously using EXOFASTv2. That study reports best-fit atmospheric parameters of $T_{\mathrm{eff}} = 5920 \pm 210$ K, $\log g = 3.86^{+0.02}_{-0.03}$ dex, and $[\mathrm{Fe/H}] = 0.18^{+0.16}_{-0.15}$ dex.

Although these parameter sets are consistent within uncertainties, the differences are sufficient to shift the stellar evolutionary track that best reproduces the observed stellar properties, leading to non negligible variations in the inferred stellar mass and age.

Given that the stellar mass inferred by \citet{vowell2025} relies on a combination of different data and modeling, it may differ slightly from the mass implied by the spectroscopic atmospheric parameters alone. 

Other relevant parameters for our MESA models are the mixing length parameter defining convection, $\alpha_{MLT}=1.73$, the overshooting parameter of $f=0.016$, and a Reimers mass-loss formalism with an efficiency of $0.1$.

We also include a simplified model of planet engulfment, shown in the bottom panel of Fig.~\ref{fig:HRage}. Rather than modeling the engulfment process self-consistently, we adopt a post-processing approach in which the accreted material is instantaneously mixed within the stellar convective envelope. The subsequent evolution of the surface lithium abundance accounts for both dilution and lithium destruction under convective conditions.

\section{Planet mass and stability limits} \label{app:pl_params}

The orbital period of the inner planet is chosen such that its initial separation from the BD spans both dynamically stable and unstable configurations. To quantify this, we characterize the system in terms of the mutual Hill radius $R_{\rm mH}$, which measures the distance at which the gravitational spheres of influence of two neighboring planets overlap. The orbital spacing between adjacent planets is commonly expressed by the dimensionless separation $\Delta_{\rm H}$ measured in units of $R_{\rm mH}$.
When planets are closely spaced in units of $R_{\rm mH}$, mutual interactions become significant \citep{1993Icar..106..247G}, and sufficiently compact configurations are expected to become dynamically unstable on long timescales. For separations  $\Delta_{\rm H}<2\sqrt{3}$, repeated encounters generally lead to rapid destabilization, representing an absolute lower limit for stability \citep{1996Icar..119..261C}. 

More recent numerical studies have shown that long-term stability in tightly packed exoplanetary systems typically requires larger separations, $\Delta_{\rm H}\sim 8-10$ \citep{2017Icar..293...52O}. The vertical blue dashed line in Figure \ref{fig:sabtility} therefore marks the commonly adopted stability threshold in mutual Hill-radius separation between the fixed BD ($m=22\,m_{\rm Jup}, P_{\rm BD}=7.1\,$d) and a hypothetical inner planet of varying mass orbiting a $M_\star=1.3\,M_\odot$ star. Inner planets 
with separations below this threshold are expected to be Hill unstable, whereas larger separations correspond to formally Hill-stable configurations, although long-term instability may still arise.

However, given the compact architecture of the TOI-5882 system, the available orbital parameter space does not allow for an inner planet to be placed at separations as large as $\Delta_H \sim 8-10$ since it would require orbital periods so short that the planet would be located unphysically close to the host star, leaving no dynamically viable region for long-term survival. We additionally mark a more conservative stability criterion at $\Delta_{\rm H} = 5$ (green dashed line) which better reflects the limited dynamical room available for an additional inner planet in this system.

The other critical parameter in the dynamical evolution of the system are the assumed properties of the accreted planet, particularly its mass. At the same time, its composition will affect the lithium enhancement of the star.

One of the main uncertainties in modeling the engulfment
concerns the mass of the accreted body, which depends sensitively on the assumed lithium abundance of the engulfed planet. \cite{2026arXiv260521407K} consider three possible compositions for the accreted material. Assuming a proto-stellar composition, they estimate an accreted mass of $\sim5.6\,m_{\rm{Jup}}$. However, as discussed in that work, such a composition is difficult to reconcile with current planet formation theories. Alternatively, adopting a higher metal content, and therefore a larger lithium reservoir, the inferred mass spans the range $9\text{-}95\,m_\oplus$. Moreover, planets in this latter mass range, in contrast to those with masses larger than $1\,m_{\rm Jup}$ are expected to deposit its material into the stellar convective envelope, and thus could reproduce the observed lithium enhancement. Accordingly, in our simulations we restrict the planetary masses to this range and do not consider the $\sim5.6\,m_{\rm{Jup}}$ scenario.

Given the presence of the observed BD with an orbital period of 7.1 days, any hypothetical inner planet would remain dynamically stable only if it were sufficiently separated from the brown dwarf. Since stability criteria depend on planetary masses \citep{1996Icar..119..261C, 2015CeMDA.123..453R}, the inclusion of an additional giant planet leaves only a very limited parameter space compatible with long-term stability. 

Although the Hill stability criterion provides a sufficient condition for stability, it is not a necessary one. Systems that formally violate the Hill criterion may nonetheless remain stable for several Gyr \citep{1980AJ.....85.1122W,1993Icar..106..247G}. However, as shown by \citet{2013ApJ...774..129D}, initially circular configurations such as those considered in our simulations that fail the Hill criterion are very likely to evolve chaotically.

While the red dashed line in Fig. \ref{fig:sabtility} defines the lower boundary for global orbital instability, the Hill stability limit marks the transition beyond which motion is guaranteed to remain bounded and non-crossing. Between these two limits lies a complex region of phase space where both long-lived and unstable configurations may coexist. Planet–planet interactions may trigger scattering events that blur this boundary: as the orbital separation increases, scattering becomes less frequent and long-term stability is more easily maintained \citep{2008ApJ...686..580C, 2017Icar..293...52O}. Long-term stability begins to emerge at separations of $\gtrsim 5\, \Delta$, as indicated by the green dashed line and consistent with the simulations in this region (dark grey crosses), which survive for nearly 5 Gyr. We also note that these stability limits account only for gravitational interactions between the planets, stellar evolution—even if its effect is small during the evolutionary stage considered here—further contribute to the system's destabilization. Since these limits are derived using the currently observed parameters of the BD, they do not necessarily represent the pre-evolutionary stability boundaries. Stellar mass loss modifies the architecture of planetary systems, such that configurations marginally stable during the main sequence may later become unstable during post-MS evolution \citep{2002ApJ...572..556D}.

\end{appendix}

\end{document}